\documentclass[aps,pra,notitlepage,twocolumn,10pt,superscriptaddress,floatfix,noeprint]{revtex4-2}
\pdfoutput=1
\usepackage[utf8]{inputenc}
\usepackage[english]{babel}
\usepackage[T1]{fontenc}
\usepackage{amsmath}
\usepackage{amsfonts}
\usepackage{amssymb}
\usepackage{amsthm}
\usepackage{makeidx}
\usepackage{graphicx}
\usepackage{color}
\usepackage{xspace}
\usepackage{enumitem}
\usepackage{hyperref}
\definecolor{mpGray}{RGB}{236, 239, 241}			
\definecolor{mpBlue}{RGB}{21, 101, 192}			
\definecolor{mpLightBlue}{RGB}{144, 202, 249}		
\definecolor{mpRed}{RGB}{198, 40, 40}			
\definecolor{mpLightRed}{RGB}{239, 154, 154}		
\definecolor{mpGreen}{RGB}{46, 125, 50}			
\definecolor{mpLightGreen}{RGB}{165, 214, 167}	
\hypersetup{citecolor=mpGreen}
\hypersetup{colorlinks=true}
\hypersetup{linkcolor=mpBlue}
\hypersetup{urlcolor=mpBlue}
\renewcommand{\paragraph}[1]{\addcontentsline{toc}{section}{#1}\emph{#1.}---}
\DeclareMathOperator{\dd}{d \!}

\DeclareMathOperator{\Ha}{\mathcal{H}}

\newcommand*{\abs}[1]{\left\lvert #1 \right\rvert}

\newcommand*{\bra}[1]{\langle #1 |}
\makeatletter
\newcommand*{\ket}[1]{| #1 \rangle \@ifnextchar\bra{\!}{}}
\makeatother

\DeclareMathOperator{\Prob}{Prob}
\DeclareMathOperator{\NN}{\mathbb{N}}
\DeclareMathOperator{\RR}{\mathbb{R}}
\DeclareMathOperator{\CC}{\mathbb{C}}
\newcommand{\nx}{X}
\newcommand{\na}{A}

\begin{document}
\title{Tsirelson inequalities: Detecting cheating and quantumness in a single framework}

\author{Martin Pl\'{a}vala}
\affiliation{Naturwissenschaftlich-Technische Fakult\"{a}t, Universit\"{a}t Siegen, Walter-Flex-Stra\ss e 3, 57068 Siegen, Germany}

\author{Teiko Heinosaari}
\affiliation{Faculty of Information Technology, Univeristy of Jyväskylä, Finland}

\author{Stefan Nimmrichter}
\affiliation{Naturwissenschaftlich-Technische Fakult\"{a}t, Universit\"{a}t Siegen, Walter-Flex-Stra\ss e 3, 57068 Siegen, Germany}

\author{Otfried G\"{u}hne}
\affiliation{Naturwissenschaftlich-Technische Fakult\"{a}t, Universit\"{a}t Siegen, Walter-Flex-Stra\ss e 3, 57068 Siegen, Germany}

\begin{abstract}
Quantumness refers to the peculiar and counterintuitive characteristics exhibited by quantum systems. Tsirelson inequalities have emerged as a powerful tool in quantum theory to detect quantumness and entanglement of harmonic oscillators, spins undergoing uniform precession, and anharmonic systems. In this paper we harness the versatility of Tsirelson inequalities to address two distinct problems: detecting cheating in classic shell games and probing quantumness in spatially separated systems and harmonic oscillators. By adopting a black-box approach and a geometric characterization of the space of conditional probabilities, we demonstrate that Tsirelson inequalities can be used in both scenarios, enabling us to uncover quantum signatures and identify cheaters in a single unified framework. This connection provides an intuitive different perspective on quantumness of mechanical systems.
\end{abstract}

\maketitle

\section{Introduction}
The term 'quantumness' refers to the unique and counterintuitive features that arise in the quantum world and distinguish quantum systems from systems obeying classical physics. Detecting quantumness has become an utmostly important task as all exciting applications of quantum technology eventually are based on some side of quantumness. There are several methods for detecting quantumness of physical systems: the most notable ones are Bell nonlocality \cite{Bell-ineq,BrunnerCavalcantiPironioScaraniWehner-BellNonlocality,RossetBancalGisin-BellNonlocality}, contextuality \cite{KochenSpecker-contextuality,Spekkens-contextuality,HobanCampbellLoukopoulosBrowne-MBQCcontextuality,Raussendorf-MBQCcontextuality,CabelloSeveriniWinter-contextuality,AmaralCuhna-contextuality,TavakoliUola-contextuality,BudroniCabelloGuhneKleinmann-contextuality,SchmidSelbyWolfeKunjwalSpekkens-noncontextuality} and schemes based on macrorealism and sequential measurements \cite{LeggettGarg-sequential,EmaryLambertNori-sequential,ClementeKofler-sequential,UolaVitaglianoBudroni-sequential}. For continuous variable systems, quantumness is often associated with the negativity of the Wigner function \cite{Wigner-WignerFunctions}, since negative Wigner functions have no classical analog. Wigner negativity is necessary for the speedup of quantum computation \cite{MariEisert-WignerFunctionsComputation,HowardWallmanVeitchEmerson-ComputationContextuality}, but also for tunneling \cite{LinDahlsten-tunneling} and incompatibility of measurements \cite{GhaiSharmaGhosh-WignerFunctionsIncomaptibility,OhstPlavala-WignerFunctionsSymmetries} and can be seen as measure of nonclassicality \cite{TanChoiJeong-WignerFunctionsNonclassicality}. Strikingly, this nonclassicality was recently demonstrated with nanomechanical oscillators of masses reaching up to micrograms, employing a tomographic scheme to reconstruct the Wigner function \cite{Chu2018,Satzinger18,Wollack22,von2022parity,bild2022}.

An interesting approach for characterizing quantumness of harmonic oscillators was initiated by Boris Tsirelson in 2006 \cite{Tsirelson-oscillatorIneq}. Consider a classical one-dimensional oscillator with period $T$, which is observed at the three time steps, $T/3$, $2T/3$, and $T$. Then, independent of the initial conditions, the position coordinate for at least one of the time steps must be negative. Consequently, the average probability of observing a positive position coordinate is bounded between ${1}/{3}$ and ${2}/{3}$; we will refer to this bound and its generalizations as Tsirelson's inequality. This, however, does not hold for quantum mechanical oscillators. If the Wigner function is positive, then Tsirelson's inequality holds, so one may see the inequality as an elegant way to detect the negativity of Wigner function, complementing other Wigner negativity witnesses \cite{FilipMista-WignerFunctionsDetectingPositive,ChabaudEmeriauGrosshans-WignerFunctionNegativeWitness}. The same approach was recently used to certify non-classicality of spin systems \cite{ZawCenxinLasmarScarani-spinTsirelsonIneq} and anharmonic systems \cite{ZawScarani-anharmonicTsirelsonIneq}, but also to detect entanglement between harmonic oscillators \cite{JayachandranZawScarani-entanglementTsirelsonIneq}.

In this paper, we formulate the aforementioned inequalities in a black-box approach in order to construct a framework in which a complete list of relevant inequalities can be explicitly derived from the underlying physical principles using convex geometry. We thus replace previous counting arguments used to guess some inequalities by a systematic approach that finds all possible inequalities. This has two important benefits: First, we are able to identify other applications of the inequalities beyond the detection of quantumness in mechanical systems; explicitly, we demonstrate the detection of cheating in the shell game and an alternative derivation of Bell inequalities. The second advantage is that we are able to find all tight inequalities for the given scenario, which we demonstrate by finding new inequalities for the quantum harmonic oscillator, some of which are violated for lower energies than previously known inequalities.

\section{Black-box approach to Tsirelson inequalities}
Consider a generic experiment in which a certain discrete event or outcome $a \in \{1,\ldots,\na\}$ is observed, conditioned on a made choice $x \in \{ 1,\ldots, \nx\}$. The conditional probabilities $[p(a|x)]_{a,x}$ determining the likelihoods of outcomes can be estimated through many trials of this experiment. The probabilities are normalized as $\sum_{a=1}^{\na} p(a|x) = 1$ for every $x$. Here we will use $[p(a|x)]_{a,x}$ to denote the whole matrix of numbers, while $p(a|x)$ will be only one entry of this matrix, i.e., a number between 0 and 1. We will investigate the question whether there exists a constrained 'classical' model for $[p(a|x)]_{a,x}$, that is, a global probability distribution $[p(a_1 \ldots a_{\nx})]_{a_1,\ldots,a_{\nx}}$ such that $[p(a|x)]_{a,x}$ can be obtained from $[p(a_1 \ldots a_{\nx})]_{a_1,\ldots,a_{\nx}}$ by marginalization, 
\begin{equation}
p(a|x) = \sum_{a_1, \ldots, a_{x-1}, a_{x+1}, \ldots, a_{\nx}} p(a_1 \ldots a_{\nx}),
\end{equation}
and such that $[p(a_1 \ldots a_{\nx})]_{a_1,\ldots,a_{\nx}}$ satisfies one or several constraints of the form: $p(\tilde{a}_1, \ldots, \tilde{a}_{\nx}) = 0$ for a specified set of indices $\tilde{a}_1, \ldots, \tilde{a}_{\nx}$. Such constraints are strictly different from the constraints in Bell and Kochen-Specker scnearios where the constraints come from the structure of measurements and from the properties of the classical model that is attempting to explain the observed results. In our case, there is no generally valid way to justify the existence of specific, here called classical, constraints; they must be derived from the structure of the specific problem at hand. We remark that constraints are necessary since without any constraints, we can set $p(a_1 \ldots a_{\nx}) = \prod_{x=1}^{\nx} p(a_x|x)$ as the global probability distribution, which has the correct marginals.

To define Tsirelson inequalities, we need to fix the size of the conditional probability matrices (i.e. the numbers $\na$ and $\nx$) and choose a set of constraints. We define a Tsirelson inequality as a linear inequality that is satisfied by all conditional probability matrices $[p(a|x)]_{a,x}$ for which a constrained classical model exists, but that is violated by some conditional probability matrix $[p(a|x)]_{a,x}$ for which a constrained classical model does not exist. That is, Tsirelson inequalities are linear witnesses that certify that for some conditional probability matrix $[p(a|x)]_{a,x}$ the classical constrained model cannot exist.

Given a set of constraints, one can find all Tsirelson inequalities as follows. First, we observe that the extreme points of the set of global probability distributions $[p(a_1 \ldots a_{\nx})]_{a_1,\ldots,a_{\nx}}$ are exactly the distributions which take the value $1$ for some specific choice of the indices $a_1 \ldots a_{\nx}$, but are zero everywhere else. By construction, these extreme points are also extremal on the convex set of the corresponding marginal distributions, since they fulfill $p(a|x)=1$ whenever $a$ matches the chosen $a_x$ for each $x$ and are zero otherwise. We can thus identify each extremal distribution by a vertex spanning the polytope of all possible probability matrices.

\begin{figure}
\includegraphics[width=0.8 \linewidth]{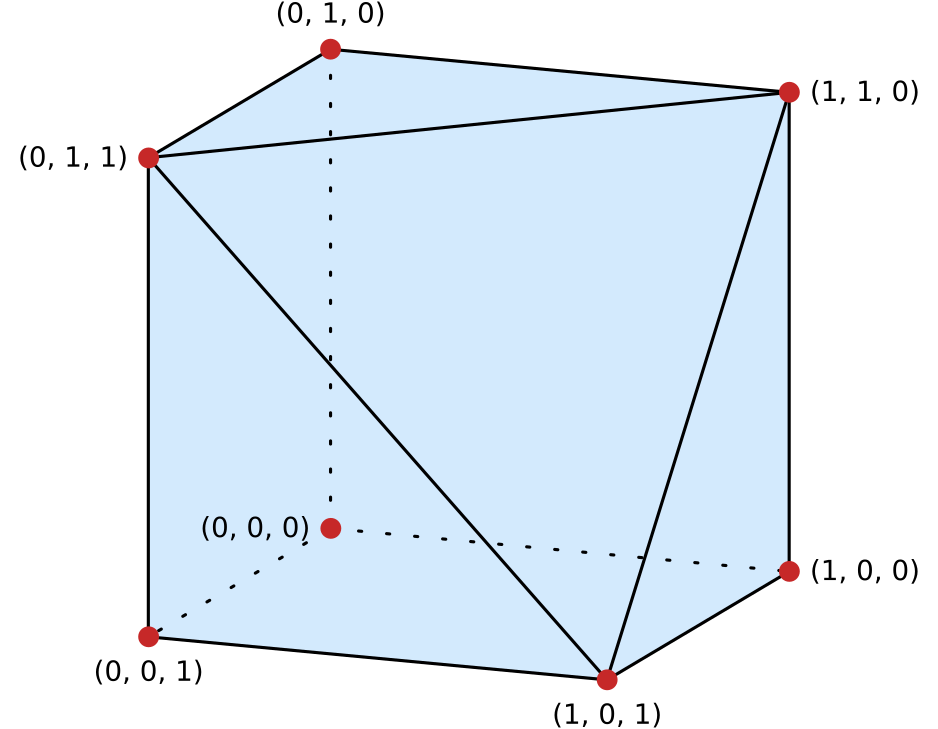}
\caption{The polytope of all conditional probability matrices $[p(a|x)]_{a,x}$ that have a constrained classical model with the constraint $p(111) = 0$. Here a matrix $[p(a|x)]_{a,x}$ is identified with a vector $(p(1|1), p(1|2), p(1|3))$.\label{fig:n3-000}}
\end{figure}

The constraints rule out some of the extreme points, but the other extreme points satisfying the constraints remain unchanged; this follows since the extreme points of the set of all probability distributions are Dirac delta distributions concentrated at single points. Accordingly, one must exclude the corresponding vertices in the polytope of conditional probability matrices $[p(a|x)]_{a,x}$, leading to a polytope $P$ of all classically constrained conditional probability matrices. New extreme points cannot arise in the space of conditional probability distributions due to linearity of the marginalization and a simple counting argument: the resulting polytope $P$ in the space of conditional probabilities can have at most the same number of extreme points as the set of probability distributions satisfying the constraints; the result follows by counting the numbers of extreme points. It also follows that whenever there is at least one constraint, then there is some conditional probability distribution $[p(a|x)]_{a,x}$ which does not have a constrained classical model.

Finally, in order to find all Tsirelson inequalities, one needs to find all facets of $P$ and express them via corresponding inequalities. This can be done, e.g., with suitable software tools such as \texttt{polymake} \cite{polymake}. While some of these inequalities will be trivial (e.g., enforce that probabilities are positive numbers), one can identify all non-trivial inequalities that will exactly correspond to Tsirelson inequalities for the given scenario.

Consider the simplest case with $\na = \nx = 2$ and only one constraint, $p(11) = 0$, which rules out one extreme point of the polytope $P$ of conditional probabilities. The remaining three extreme points of $P$ form a triangle, and the only non-trivial Tsirelson inequality is $p(1|1) + p(1|2) \leq 1$.

\begin{figure}
\includegraphics[width=0.8 \linewidth]{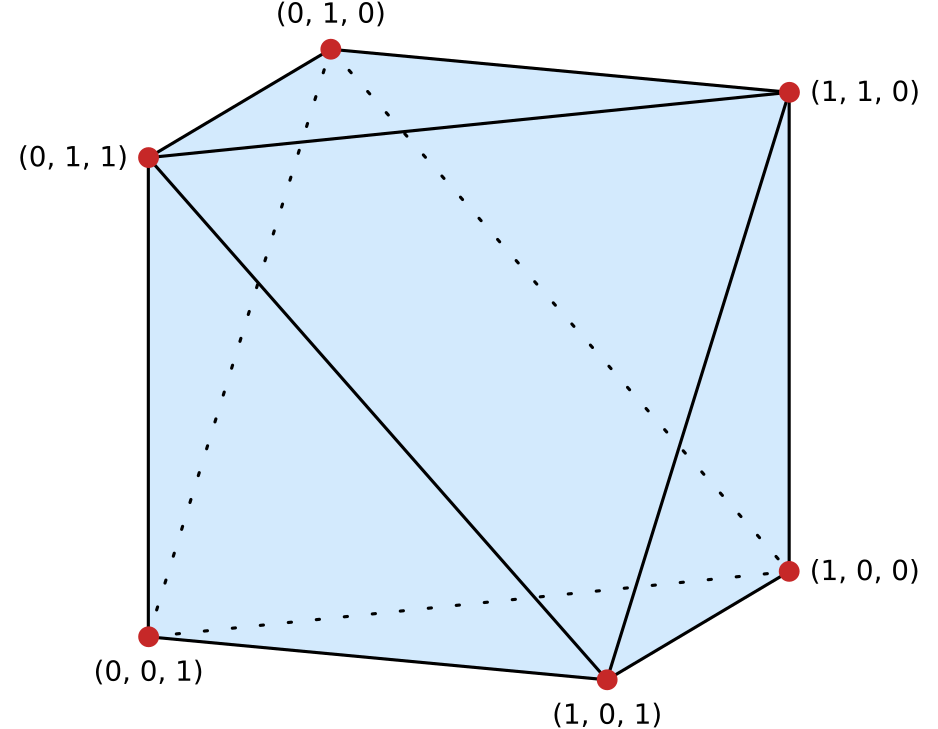}
\caption{The polytope of all conditional probability matrices $[p(a|x)]_{a,x}$ that have a constrained classical model with the constraints $p(000) = 0$ and $p(111) = 0$. A matrix $[p(a|x)]_{a,x}$ is again identified with a vector $(p(1|1), p(1|2), p(1|3))$ \label{fig:n3-Tsirelson}}
\end{figure}

The situation becomes more interesting already for $\nx = 3$ and $\na = 2$. Let us consider the constraint $p(111) = 0$ ruling out the vertex given by $p(1|x) = 1$ for all $x$. The set of conditional probabilities with constrained classical models has $7$ extreme points; see Fig~\ref{fig:n3-000}, and the only Tsirelson inequality in this case is
\begin{equation} \label{eq:n3-000}
p(1|1) + p(1|2) + p(1|3) \leq 2 \, .
\end{equation}
As we will discuss later, Tsirelson \cite{Tsirelson-oscillatorIneq} considered the two constraints $p(111) = 0$ and $p(222) = 0$. Then the set of conditional probabilities with constrained classical models has $6$ extreme points: the same as in the previous case except for the one given by $p(1|x) = 0$ for all $x$, see Fig.~\ref{fig:n3-Tsirelson}. The Tsirelson inequalities in this case are
\begin{equation} \label{eq:n3-Tsirelson}
1 \leq p(1|1) + p(1|2) + p(1|3) \leq 2.
\end{equation}

\section{Non-linear constraints}
More generally, one can consider constraints of the form $\lambda_{\min} \leq f([p(a_1 \ldots a_\nx)]_{a_1, \ldots, a_\nx}) \leq \lambda_{\max}$ for some $\lambda_{\min}, \lambda_{\max} \in \RR$. If the function $f$ is linear, the described methods generalize straightforwardly, and one also obtains a polytope $P$ of classically constrained conditional probabilities, whose facets yield the sought inequalities. If $f$ is nonlinear, then the set of constrained classical probabilities and their marginals may not be a polytope, in which case the set of all Tsirelson inequalities is not finite and cannot be found as in the case of linear constraints. Nevertheless one may try to find nonlinear inequalities based on entropy \cite{augusiak2009positive}, but no general procedure is known that would be universally applicable. To demonstrate this we present a simple example of non-linear constraints.

\begin{figure}
\includegraphics[width=\linewidth]{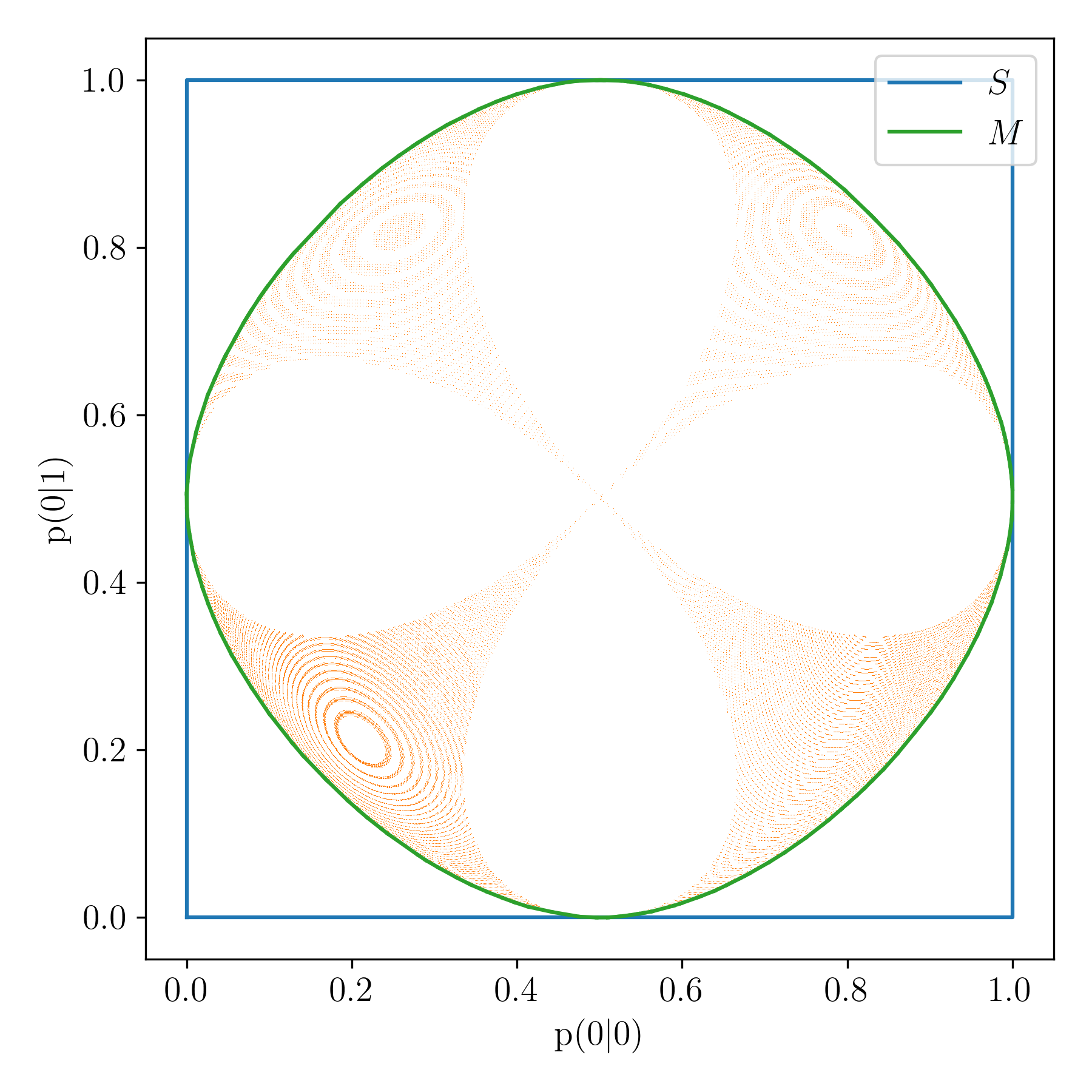}
\caption{The polytope $S$ of all conditional probability matrices $[p(a|x)]_{a,x}$ and the convex body $M$ of all conditional probabilities with a constrained 'classical' model. The orange dots are the marginals of the sampled extreme points of the set of 'classical' probability distributions satisfying the constraint \eqref{eq:nonlinear-constraint}, they do not fill the entire interior of $M$ since they correspond on to the surface of the 'classical' probability distributions satisfying the constraint \eqref{eq:nonlinear-constraint}. Here a matrix $[p(a|x)]_{a,x}$ is identified with a vector $(p(1|1), p(1|2))$ and $N = 500$ was used for the sampling.\label{fig:nonlinear}}
\end{figure}

Let $\na = \nx = 2$ and consider the non-linear constraint
\begin{equation} \label{eq:nonlinear-constraint}
\sum_{a_1, a_2 = 1}^2 \left( p(a_1 a_2) - \frac{1}{4} \right)^2 \leq \frac{1}{4}.
\end{equation}
This constraint enforces that the probability vector of the 'classical' model is contained in the ball of radius $\frac{1}{2}$ centered at the uniform distribution given by $p(a_1 a_2) = \frac{1}{4}$ for all $a_1, a_2 \in \{1,2\}$. In this case, one can numerically compute the set of all conditional probabilities $[p(a|x)]_{a,x}$ that have the constrained classical model as follows: we can iterate over subset vectors $\vec{x} \in \RR^4$ such that $\sum_{i=1}^4 (x_i - \frac{1}{4})^2 = \frac{1}{4}$, i.e., that belong to the surface of the ball corresponding to the non-linear constraint \eqref{eq:nonlinear-constraint}. We can do this by considering the parametrization
\begin{equation} \label{eq:nonlinear-xi}
\begin{split}
x_1 &= \frac{1}{2} \cos(\varphi_1) + \frac{1}{4} \\
x_2 &= \frac{1}{2} \sin(\varphi_1) \cos(\varphi_2) + \frac{1}{4} \\
x_3 &= \frac{1}{2} \sin(\varphi_1) \sin(\varphi_2) \cos(\varphi_3) + \frac{1}{4} \\
x_4 &= \frac{1}{2} \sin(\varphi_1) \sin(\varphi_2) \sin(\varphi_3) + \frac{1}{4}
\end{split}
\end{equation}
and by picking $\varphi_1 = \frac{k_1}{N} \pi$, $\varphi_1 = \frac{k_1}{N} \pi$, $\varphi_1 = \frac{k_3}{2N} 2\pi$ for $k_1, k_2 \in \{1, \ldots, N\}$ and $k_3 \in \{1, \ldots, 2N\}$ and $N \in \NN$. Then, if for such vector we have $\sum_{i=1}^4 x_i = 1$ and $x_i \geq 0$ for all $i \in \{1, \ldots, 4\}$, $\vec{x}$ corresponds to an extreme point of the set of 'classical' probability distributions satisfying the constraint \eqref{eq:nonlinear-constraint}. Then we just compute the corresponding conditional probability $[p(a|x)]_{a,x}$ by marginalization and subsequently, we can find the extreme points of this approximation of the set of all conditional probabilities with constrained 'classical' models. The results are plotted in Fig.~\ref{fig:nonlinear}.

\section{Shell game}
The shell game is a game famously played on the streets of New York and other big cities. The dealer presents three cups and one ball, the ball is hidden beneath one of the cups, and the cups are shuffled. The player's task is to determine under which cup the ball is hidden. The usual catch is that the dealer is cheating by removing the ball, in which case the game is unwinnable for the player. The dealer can, of course, also add more balls to the game, making the game easier for the player.

We interpret the conditional probability $p(2|x)$ as the probability of finding the ball if the cup labeled by $x$ is chosen, thus we have $\na = 2$ and $\nx = 3$. The classical model $p(a_1 a_2 a_3)$ will be interpreted as follows: $p(111)$ is the probability that there is no ball under any cup, $p(211)$, $p(121)$, $p(112)$ is the probability that the ball is under the first, second, third cup, respectively, while e.g. $p(221)$ is the probability that there are balls under the first and second cup.  Finally, $p(222)$ is the probability that there is a ball under every cup. The constraint corresponds to our assumption that there is at least one ball in the game, which means that we must have $p(111) = 0$. It follows that the Tsirelson inequality \eqref{eq:n3-000} must be satisfied and, using $p(2|x) = 1 - p(1|x)$, we can write
\begin{equation}
\dfrac{1}{3} \leq \dfrac{\sum_{x=1}^3 p(2|x)}{3} \, .
\end{equation}
The quantity on the right-hand-side is nothing else than the average probability of winning if we choose a random cup. We thus see that, in this case, a violation of the Tsirelson inequality \eqref{eq:n3-000} means that the dealer must be cheating by removing the ball from the game at some rounds. Here we have considered that the shell game is played with classical systems, but it is also possible to violate the inequality \eqref{eq:n3-000} with quantum systems. One may thus consider a quantum version of the shell game which might give additional advantage to the dealer; this would be similar but not directly related to the Specker’s parable of the overprotective seer \cite{LiangSpekkensWiseman-Specker,LiangSpekkensWiseman-SpeckerErratum}.

\section{Bell inequalities}
To demonstrate that our formalism can be used to derive Bell inequalities \cite{Bell-ineq, BrunnerCavalcantiPironioScaraniWehner-BellNonlocality,RossetBancalGisin-BellNonlocality}, we now show the derivation of the CHSH \cite{ClauserHorneShimonyHolt-CHSH,ClauserHorneShimonyHolt-CHSHerratum} inequality, leaving generalizations for future work. Our approach is reminiscent of previous approaches to contextuality \cite{AbramskyBrandenburger-contextuality,AbramskyHardy-logicalBell}. In the CHSH scenario, we have two spatially separated parties, Alice and Bob, each of which can choose one of two dichotomic measurements, $x,y \in \{1, 2\}$. Hence we have $X=4$, and we represent the choices using $xy \in \{11, 12, 21, 22\}$. Let us group the four possible outcomes Alice and Bob can measure by $A=2$ distinct results: $p(1|xy)$ will denote the probability that Alice and Bob obtain the same outcome if Alice chooses the measurement $x$ and Bob chooses $y$, whereas $p(2|xy)$ corresponds to obtaining different outcomes. The classical model $p(a_{11} a_{12} a_{21} a_{22})$ will be represented as follows: $p(1111)$ is the probability that Alice and Bob always obtain the same outcome, $p(1112)$ is the probability that Alice and Bob obtain the same outcome if $x+y \leq 3$, but the opposite outcome if $x = y = 2$, and analogically for the other options.

Now assume that the experiment is classical and we can measure all of the $xy$-possibilities simultaneously. Then $p(1112) = p(2221) = 0$, because these options violate basic logic: In the $p(1112)$ case, Alice and Bob always must get the same outcome for $xy=11,12,21$, and different outcomes for $xy=22$. If both Alice and Bob choose the first measurement ($x=y=1$) and Alice obtains, say, $1$, then also Bob obtains $1$. It follows that Bob obtains $1$ also if $x=1$ and $y=2$, and so Bob always obtains $1$. If $x=2$, $y=1$ are chosen, then Alice must obtain $1$, because her outcome must be the same as Bob's and Bob always gets $1$. But then they would always get both $1$ also for the $x=y=2$ choice, which is a contradiction. The $p(2221)$ case follows similarly. From these constraints and the previously outlined methods, we get the inequalities $0 \leq p(1|11) + p(1|12) + p(1|21) - p(1|22) \leq 2$. Expressing $p(1|xy) = \frac{1}{2}(1 + E_{xy})$ in terms of the correlation $E_{xy}$, we arrive at $-2 \leq E_{00} + E_{01} + E_{10} - E_{11} \leq 2$.

\section{Quantum harmonic oscillators}
Let us now analyze how generalized Tsirelson inequalities are used to certify the non-classicality of states of a harmonic oscillator of period $T$. Given some initial state, we fix $\nx \in \NN$ and measure the position of the oscillator at time $t_x = \frac{x}{\nx} T$, for a chosen $x \in \{1, \ldots, \nx\}$. We repeat this prepare-measure procedure many times (with the same initial state) in order to obtain measurement data at every possible time $t_x$. Since each position measurement is made on a newly initialized oscillator state, the measurement backaction plays no role, unlike in a Leggett-Garg test scheme \cite{LeggettGarg-sequential,EmaryLambertNori-sequential,UolaVitaglianoBudroni-sequential}.

We are only interested whether the measured position is positive or negative; we will use $\Prob(q > 0| t_x)$, $\Prob(q = 0| t_x)$ to denote the probability that the position measured at time $t_x$ is strictly positive, or equal to zero, respectively. Then for $\na = 2$ we set
\begin{equation} \label{eq:HO-condProb}
p(1|x) = \Prob(q > 0| t_x) + \dfrac{1}{2} \Prob(q = 0| t_x),
\end{equation}
where the latter term is added for consistency, see the explanation in \cite{ZawCenxinLasmarScarani-spinTsirelsonIneq}. We will show that if the initial state of the harmonic oscillator is classical then the conditional probability matrix $[p(a|x)]_{a,x}$ has a constrained classical model. This result will enable us to derive inequalities that certify the non-classicality of the initial state.

Given a wave function $\ket{\psi} \in \Ha$, where $\Ha$ is the Hilbert space corresponding to single harmonic oscillator, we define the Wigner function $W_\psi$ corresponding to $\ket{\psi}$ as
\begin{equation}
W_\psi(q,p) = \dfrac{1}{\pi \hbar} \int_{\RR} \bar{\psi}(q + x) \psi(q - x) e^{i\frac{2px}{\hbar}} \dd x \, .
\end{equation}
It represents the wave function in phase space, with position and momentum coordinates $q$ and $p$. The Wigner-Weyl picture in which Wigner functions replace wave functions is an equivalent description of quantum theory \cite{Wigner-WignerFunctions,Groenewold-QM,Moyal-WignerFunctions,CurtrightFarlieZachos-WignerFunctions}. The probability distribution of position is obtained by marginalization of $W_\psi$,
\begin{equation} \label{eq:HO-ProbQ}
\Prob(q) = \abs{\psi(q)}^2 = \int_{\RR} W_\psi(q,p) \dd p.
\end{equation}
A state of a classical harmonic oscillator is given by a probability density in phase space $\rho(q,p)$, where we now require that $\rho(q,p) \geq 0$ for all $q,p$. For example, the state of a localized particle is described by a Dirac delta distribution. It is well-known that for the harmonic oscillator, the time evolution of classical and quantum harmonic oscillator in phase space coincides \cite{Case-wignerFunctions}; and the formula \eqref{eq:HO-ProbQ} for the position marginal coincides, too. Thus the only non-classical property that the quantum harmonic oscillator can demonstrate in this experiment is the negativity of the Wigner function $W_\psi$ of the initial state.

Now assume that the Wigner function of the initial state is positive, $W_\psi(q,p) \geq 0$ for all $q,p$. Then we can formally treat $W_\psi$ as a state of a classical harmonic oscillator, and we can thus, in principle, assume that we can measure the position of the oscillator without disturbing its state. Instead of measuring position only at one fixed time $t_x$, we could measure it at every possible time $t_x$ in a single run of the experiment. In this way, we would obtain the global probability distribution $[p(a_1 \ldots a_{\nx})]_{a_1,\ldots,a_{\nx}}$, the marginals of which are the $[p(a|x)]_{a,x}$ given in \eqref{eq:HO-condProb}.

Unless it is in the stationary ground state, a harmonic oscillator swings periodically in configuration space and the position of the oscillator is continuously positive for half of the period and continuously negative for the other half. This basic insight, which constrains the probability distribution $[p(a_1 \ldots a_{\nx})]_{a_1,\ldots,a_{\nx}}$ corresponding to a Dirac delta distribution in phase space, can be used to obtain Tsirelson inequalities, since every positive Wigner function can be expressed as a convex combination of Dirac delta distributions as $W_\psi(q,p) = \iint_{\RR^2} W_\psi(\tilde{q},\tilde{p}) \delta(p-\tilde{p}) \delta(q-\tilde{q}) \dd \tilde{q} \dd \tilde{p}$. In the following, we will investigate only the cases where $\nx = 3$ and $\nx = 5$, but a similar analysis can be carried out for arbitrary $\nx$.

For $\nx = 3$, we have to observe the position to be positive at least once, but at most twice, since the oscillator must eventually swing the other way. We thus get the classical constraints $p(111) = 0$, which means that the position cannot be always positive, and $p(222) = 0$, which means that the position cannot be always negative. This is exactly the case originally analyzed by Tsirelson \cite{Tsirelson-oscillatorIneq}, and there are only the two Tsirelson inequalities \eqref{eq:n3-Tsirelson}.

\begin{table}
\caption{\label{tab:HO-nx5}Maximal violations of the original Tsirelson inequalities \eqref{eq:HO-TsirelsonIneq} (denoted as type T), type I inequalities \eqref{eq:HO-ineq1}, and type II inequalities \eqref{eq:HO-ineq2} by oscillator states that are superpositions of the first $N$ eigenstates of the Hamiltonian. Here we compute only the violations of the respective upper bounds, one obtains analogical results for lower bounds. Numerical values are rounded to three decimal places, blank space indicates no violation for the given $N$.}
\begin{ruledtabular}
\begin{tabular}{cccc}
$N$ & Type T & Type I & Type II\\
\hline
5 & & 2.007 & \\
6 & 3.046 & 2.01 & 1.018 \\
10 & 3.046 & 2.038 & 1.046 \\
15 & 3.11 & 2.067 & 1.072 \\
\end{tabular}
\end{ruledtabular}
\end{table}

For $\nx = 5$, we have to observe the position to be positive at least twice, but at most thrice, because the times of observation $t_x$ are evenly spread over the single period $T$. We thus get the constraints $p(11111) = 0$, $p(22222) = 0$, and all cyclic permutations of $p(21111) = 0$ and $p(22221) = 0$. Using these constraints, we obtain
\begin{equation} \label{eq:HO-TsirelsonIneq}
2 \leq \sum_{x=1}^5 p(1|x) \leq 3,
\end{equation}
a generalization of the inequalities \eqref{eq:n3-Tsirelson} already derived in \cite{ZawCenxinLasmarScarani-spinTsirelsonIneq}. We obtain new inequalities by using additional constraints due to the fact that we observe the oscillator during a single period. We must have $p(21212) = 0$, $p(12121) = 0$, and all cyclic permutations, which yields additional inequalities  we term type I and II, let $x \in \{1, \ldots, 5\}$ and $\oplus$ denote addition modulo $5$, then we have
\begin{align}
\text{type I:} \,\,\,\, & 1 \leq p(1|x) + p(1|x \oplus 2) + p(1|x \oplus 4) \leq 2, \label{eq:HO-ineq1} \\
\text{type II:} \,\,\,\, & 0 \leq p(1|x) - p(1|x \oplus 1) + p(1|x \oplus 2) \leq 1. \label{eq:HO-ineq2}
\end{align}

While it is currently unknown how to find the highest possible violation of the inequalities \eqref{eq:HO-TsirelsonIneq} - \eqref{eq:HO-ineq2}, we can look for the maximal violation by a quantum state restricted to the subspace of the first $N$ eigenstates of the Hamiltonian: $\ket{\psi} = \sum_{n=0}^{N-1} \alpha_n \ket{n}$,  where $\ket{n}$ denotes the $n$\textsuperscript{th} eigenstate and $\alpha_n \in \CC$. This can be done numerically in a similar manner as described by Tsirelson in \cite{Tsirelson-oscillatorIneq}. While the original and the type II inequalities \eqref{eq:HO-TsirelsonIneq} and \eqref{eq:HO-ineq2} are only violated for $N\geq 6$, the type I inequalities \eqref{eq:HO-ineq1} are already violated for $N=5$. This may prove viable for experimental tests of nonclassicality with oscillators. Table~\ref{tab:HO-nx5} displays the detailed numerical results. 

\section{Conclusions}
We have formulated a black-box approach to Tsirelson inequalities, which enabled us to systematically derive further inequalities corresponding to the facets of a polytope. One can apply this approach not only to construct witnesses for the negativity of the Wigner function and for the entanglement of physical systems undergoing periodic time evolution, but also, as we have shown, to detect cheating in the shell game, for example. Using our formulation based on constrained classical models, one can also look for Tsirelson inequalities for quantum backflow \cite{Allcock-backflow,Bracken-backflow}, quantum tunneling \cite{LinDahlsten-tunneling}, and other mechanical tasks \cite{Goussev-reentry,TrilloLeNavascues-rocket}, but these investigations are left for future work. It is also clear that our derivation of the CHSH inequality can be generalized to more parties and outcomes, which offers another future avenue of research.

Our findings demonstrate that Tsirelson inequalities serve as a unifying framework for modeling and hypothesis testing in diverse settings ranging from games of deception to probing the quantumness of oscillators and other systems. Just as the geometric characterization of probabilities in correlated classical or quantum systems, this may pave the way to applications in quantum foundations and technology.

\begin{acknowledgments}
We acknowledge support from the Deutsche Forschungsgemeinschaft (DFG, German Research Foundation, project numbers 447948357 and 440958198), the Sino-German Center for Research Promotion (Project M-0294), the German Ministry of Education and Research (Project QuKuK, BMBF Grant No. 16KIS1618K), the DAAD, and the Alexander von Humboldt Foundation.
TH acknowledges financial support from the Business Finland under the project TORQS, Grant 8582/31/2022, and from the Academy of Finland under the project DEQSE, Grant 349945.
\end{acknowledgments}

\bibliography{citations}

\begin{thebibliography}{50}%
\makeatletter
\providecommand \@ifxundefined [1]{%
 \@ifx{#1\undefined}
}%
\providecommand \@ifnum [1]{%
 \ifnum #1\expandafter \@firstoftwo
 \else \expandafter \@secondoftwo
 \fi
}%
\providecommand \@ifx [1]{%
 \ifx #1\expandafter \@firstoftwo
 \else \expandafter \@secondoftwo
 \fi
}%
\providecommand \natexlab [1]{#1}%
\providecommand \enquote  [1]{``#1''}%
\providecommand \bibnamefont  [1]{#1}%
\providecommand \bibfnamefont [1]{#1}%
\providecommand \citenamefont [1]{#1}%
\providecommand \href@noop [0]{\@secondoftwo}%
\providecommand \href [0]{\begingroup \@sanitize@url \@href}%
\providecommand \@href[1]{\@@startlink{#1}\@@href}%
\providecommand \@@href[1]{\endgroup#1\@@endlink}%
\providecommand \@sanitize@url [0]{\catcode `\\12\catcode `\$12\catcode
  `\&12\catcode `\#12\catcode `\^12\catcode `\_12\catcode `\%12\relax}%
\providecommand \@@startlink[1]{}%
\providecommand \@@endlink[0]{}%
\providecommand \url  [0]{\begingroup\@sanitize@url \@url }%
\providecommand \@url [1]{\endgroup\@href {#1}{\urlprefix }}%
\providecommand \urlprefix  [0]{URL }%
\providecommand \Eprint [0]{\href }%
\providecommand \doibase [0]{https://doi.org/}%
\providecommand \selectlanguage [0]{\@gobble}%
\providecommand \bibinfo  [0]{\@secondoftwo}%
\providecommand \bibfield  [0]{\@secondoftwo}%
\providecommand \translation [1]{[#1]}%
\providecommand \BibitemOpen [0]{}%
\providecommand \bibitemStop [0]{}%
\providecommand \bibitemNoStop [0]{.\EOS\space}%
\providecommand \EOS [0]{\spacefactor3000\relax}%
\providecommand \BibitemShut  [1]{\csname bibitem#1\endcsname}%
\let\auto@bib@innerbib\@empty
\bibitem [{\citenamefont {Bell}(1964)}]{Bell-ineq}%
  \BibitemOpen
  \bibfield  {author} {\bibinfo {author} {\bibfnamefont {J.~S.}\ \bibnamefont
  {Bell}},\ }\bibfield  {title} {\bibinfo {title} {{On the Einstein Podolsky
  Rosen Paradox}},\ }\href
  {https://doi.org/10.1103/PhysicsPhysiqueFizika.1.195} {\bibfield  {journal}
  {\bibinfo  {journal} {Physics}\ }\textbf {\bibinfo {volume} {1}},\ \bibinfo
  {pages} {195 } (\bibinfo {year} {1964})}\BibitemShut {NoStop}%
\bibitem [{\citenamefont {Brunner}\ \emph {et~al.}(2014)\citenamefont
  {Brunner}, \citenamefont {Cavalcanti}, \citenamefont {Pironio}, \citenamefont
  {Scarani},\ and\ \citenamefont
  {Wehner}}]{BrunnerCavalcantiPironioScaraniWehner-BellNonlocality}%
  \BibitemOpen
  \bibfield  {author} {\bibinfo {author} {\bibfnamefont {N.}~\bibnamefont
  {Brunner}}, \bibinfo {author} {\bibfnamefont {D.}~\bibnamefont {Cavalcanti}},
  \bibinfo {author} {\bibfnamefont {S.}~\bibnamefont {Pironio}}, \bibinfo
  {author} {\bibfnamefont {V.}~\bibnamefont {Scarani}},\ and\ \bibinfo {author}
  {\bibfnamefont {S.}~\bibnamefont {Wehner}},\ }\bibfield  {title} {\bibinfo
  {title} {{Bell nonlocality}},\ }\href
  {https://doi.org/10.1103/RevModPhys.86.419} {\bibfield  {journal} {\bibinfo
  {journal} {Reviews of Modern Physics}\ }\textbf {\bibinfo {volume} {86}},\
  \bibinfo {pages} {419} (\bibinfo {year} {2014})}\BibitemShut {NoStop}%
\bibitem [{\citenamefont {Rosset}\ \emph {et~al.}(2014)\citenamefont {Rosset},
  \citenamefont {Bancal},\ and\ \citenamefont
  {Gisin}}]{RossetBancalGisin-BellNonlocality}%
  \BibitemOpen
  \bibfield  {author} {\bibinfo {author} {\bibfnamefont {D.}~\bibnamefont
  {Rosset}}, \bibinfo {author} {\bibfnamefont {J.-D.}\ \bibnamefont {Bancal}},\
  and\ \bibinfo {author} {\bibfnamefont {N.}~\bibnamefont {Gisin}},\ }\bibfield
   {title} {\bibinfo {title} {{Classifying 50 years of {B}ell inequalities}},\
  }\href {https://doi.org/10.1088/1751-8113/47/42/424022} {\bibfield  {journal}
  {\bibinfo  {journal} {Journal of Physics A: Mathematical and Theoretical}\
  }\textbf {\bibinfo {volume} {47}},\ \bibinfo {pages} {424022} (\bibinfo
  {year} {2014})}\BibitemShut {NoStop}%
\bibitem [{\citenamefont {Kochen}\ and\ \citenamefont
  {Specker}(1967)}]{KochenSpecker-contextuality}%
  \BibitemOpen
  \bibfield  {author} {\bibinfo {author} {\bibfnamefont {S.}~\bibnamefont
  {Kochen}}\ and\ \bibinfo {author} {\bibfnamefont {E.}~\bibnamefont
  {Specker}},\ }\bibfield  {title} {\bibinfo {title} {The problem of hidden
  variables in quantum mechanics},\ }\href
  {https://doi.org/10.1512/iumj.1968.17.17004} {\bibfield  {journal} {\bibinfo
  {journal} {Journal of Mathematics and Mechanis}\ }\textbf {\bibinfo {volume}
  {17}},\ \bibinfo {pages} {59} (\bibinfo {year} {1967})}\BibitemShut {NoStop}%
\bibitem [{\citenamefont {Spekkens}(2005)}]{Spekkens-contextuality}%
  \BibitemOpen
  \bibfield  {author} {\bibinfo {author} {\bibfnamefont {R.~W.}\ \bibnamefont
  {Spekkens}},\ }\bibfield  {title} {\bibinfo {title} {{Contextuality for
  preparations, transformations, and unsharp measurements}},\ }\href
  {https://doi.org/10.1103/PhysRevA.71.052108} {\bibfield  {journal} {\bibinfo
  {journal} {Physical Review A}\ }\textbf {\bibinfo {volume} {71}},\ \bibinfo
  {pages} {052108} (\bibinfo {year} {2005})}\BibitemShut {NoStop}%
\bibitem [{\citenamefont {Hoban}\ \emph {et~al.}(2011)\citenamefont {Hoban},
  \citenamefont {Campbell}, \citenamefont {Loukopoulos},\ and\ \citenamefont
  {Browne}}]{HobanCampbellLoukopoulosBrowne-MBQCcontextuality}%
  \BibitemOpen
  \bibfield  {author} {\bibinfo {author} {\bibfnamefont {M.~J.}\ \bibnamefont
  {Hoban}}, \bibinfo {author} {\bibfnamefont {E.~T.}\ \bibnamefont {Campbell}},
  \bibinfo {author} {\bibfnamefont {K.}~\bibnamefont {Loukopoulos}},\ and\
  \bibinfo {author} {\bibfnamefont {D.~E.}\ \bibnamefont {Browne}},\ }\bibfield
   {title} {\bibinfo {title} {Non-adaptive measurement-based quantum
  computation and multi-party {B}ell inequalities},\ }\href
  {https://doi.org/10.1088/1367-2630/13/2/023014} {\bibfield  {journal}
  {\bibinfo  {journal} {New Journal of Physics}\ }\textbf {\bibinfo {volume}
  {13}},\ \bibinfo {pages} {023014} (\bibinfo {year} {2011})}\BibitemShut
  {NoStop}%
\bibitem [{\citenamefont {Raussendorf}(2013)}]{Raussendorf-MBQCcontextuality}%
  \BibitemOpen
  \bibfield  {author} {\bibinfo {author} {\bibfnamefont {R.}~\bibnamefont
  {Raussendorf}},\ }\bibfield  {title} {\bibinfo {title} {Contextuality in
  measurement-based quantum computation},\ }\href
  {https://doi.org/10.1103/PhysRevA.88.022322} {\bibfield  {journal} {\bibinfo
  {journal} {Physical Review A}\ }\textbf {\bibinfo {volume} {88}},\ \bibinfo
  {pages} {022322} (\bibinfo {year} {2013})}\BibitemShut {NoStop}%
\bibitem [{\citenamefont {Cabello}\ \emph {et~al.}(2014)\citenamefont
  {Cabello}, \citenamefont {Severini},\ and\ \citenamefont
  {Winter}}]{CabelloSeveriniWinter-contextuality}%
  \BibitemOpen
  \bibfield  {author} {\bibinfo {author} {\bibfnamefont {A.}~\bibnamefont
  {Cabello}}, \bibinfo {author} {\bibfnamefont {S.}~\bibnamefont {Severini}},\
  and\ \bibinfo {author} {\bibfnamefont {A.}~\bibnamefont {Winter}},\
  }\bibfield  {title} {\bibinfo {title} {Graph-theoretic approach to quantum
  correlations},\ }\href {https://doi.org/10.1103/PhysRevLett.112.040401}
  {\bibfield  {journal} {\bibinfo  {journal} {Physical Review Letters}\
  }\textbf {\bibinfo {volume} {112}},\ \bibinfo {pages} {040401} (\bibinfo
  {year} {2014})}\BibitemShut {NoStop}%
\bibitem [{\citenamefont {Amaral}\ and\ \citenamefont
  {Cunha}(2018)}]{AmaralCuhna-contextuality}%
  \BibitemOpen
  \bibfield  {author} {\bibinfo {author} {\bibfnamefont {B.}~\bibnamefont
  {Amaral}}\ and\ \bibinfo {author} {\bibfnamefont {M.}~\bibnamefont {Cunha}},\
  }\href@noop {} {\emph {\bibinfo {title} {On Graph Approaches to Contextuality
  and their Role in Quantum Theory}}},\ SpringerBriefs in Mathematics\
  (\bibinfo  {publisher} {Springer International Publishing},\ \bibinfo {year}
  {2018})\BibitemShut {NoStop}%
\bibitem [{\citenamefont {Tavakoli}\ and\ \citenamefont
  {Uola}(2020)}]{TavakoliUola-contextuality}%
  \BibitemOpen
  \bibfield  {author} {\bibinfo {author} {\bibfnamefont {A.}~\bibnamefont
  {Tavakoli}}\ and\ \bibinfo {author} {\bibfnamefont {R.}~\bibnamefont
  {Uola}},\ }\bibfield  {title} {\bibinfo {title} {{Measurement incompatibility
  and steering are necessary and sufficient for operational contextuality}},\
  }\href {https://doi.org/10.1103/PhysRevResearch.2.013011} {\bibfield
  {journal} {\bibinfo  {journal} {Physical Review Research}\ }\textbf {\bibinfo
  {volume} {2}},\ \bibinfo {pages} {013011} (\bibinfo {year}
  {2020})}\BibitemShut {NoStop}%
\bibitem [{\citenamefont {Budroni}\ \emph {et~al.}(2022)\citenamefont
  {Budroni}, \citenamefont {Cabello}, \citenamefont {Gühne}, \citenamefont
  {Kleinmann},\ and\ \citenamefont {Åke
  Larsson}}]{BudroniCabelloGuhneKleinmann-contextuality}%
  \BibitemOpen
  \bibfield  {author} {\bibinfo {author} {\bibfnamefont {C.}~\bibnamefont
  {Budroni}}, \bibinfo {author} {\bibfnamefont {A.}~\bibnamefont {Cabello}},
  \bibinfo {author} {\bibfnamefont {O.}~\bibnamefont {Gühne}}, \bibinfo
  {author} {\bibfnamefont {M.}~\bibnamefont {Kleinmann}},\ and\ \bibinfo
  {author} {\bibfnamefont {J.}~\bibnamefont {Åke Larsson}},\ }\bibfield
  {title} {\bibinfo {title} {Kochen-specker contextuality},\ }\href
  {https://doi.org/10.1103/RevModPhys.94.045007} {\bibfield  {journal}
  {\bibinfo  {journal} {Reviews of Modern Physics}\ }\textbf {\bibinfo {volume}
  {94}},\ \bibinfo {pages} {045007} (\bibinfo {year} {2022})}\BibitemShut
  {NoStop}%
\bibitem [{\citenamefont {Schmid}\ \emph {et~al.}(2021)\citenamefont {Schmid},
  \citenamefont {Selby}, \citenamefont {Wolfe}, \citenamefont {Kunjwal},\ and\
  \citenamefont {Spekkens}}]{SchmidSelbyWolfeKunjwalSpekkens-noncontextuality}%
  \BibitemOpen
  \bibfield  {author} {\bibinfo {author} {\bibfnamefont {D.}~\bibnamefont
  {Schmid}}, \bibinfo {author} {\bibfnamefont {J.~H.}\ \bibnamefont {Selby}},
  \bibinfo {author} {\bibfnamefont {E.}~\bibnamefont {Wolfe}}, \bibinfo
  {author} {\bibfnamefont {R.}~\bibnamefont {Kunjwal}},\ and\ \bibinfo {author}
  {\bibfnamefont {R.~W.}\ \bibnamefont {Spekkens}},\ }\bibfield  {title}
  {\bibinfo {title} {{Characterization of Noncontextuality in the Framework of
  Generalized Probabilistic Theories}},\ }\href
  {https://doi.org/10.1103/PRXQuantum.2.010331} {\bibfield  {journal} {\bibinfo
   {journal} {PRX Quantum}\ }\textbf {\bibinfo {volume} {2}},\ \bibinfo {pages}
  {010331} (\bibinfo {year} {2021})}\BibitemShut {NoStop}%
\bibitem [{\citenamefont {Leggett}\ and\ \citenamefont
  {Garg}(1985)}]{LeggettGarg-sequential}%
  \BibitemOpen
  \bibfield  {author} {\bibinfo {author} {\bibfnamefont {A.~J.}\ \bibnamefont
  {Leggett}}\ and\ \bibinfo {author} {\bibfnamefont {A.}~\bibnamefont {Garg}},\
  }\bibfield  {title} {\bibinfo {title} {Quantum mechanics versus macroscopic
  realism: Is the flux there when nobody looks?},\ }\href
  {https://doi.org/10.1103/PhysRevLett.54.857} {\bibfield  {journal} {\bibinfo
  {journal} {Physical Review Letters}\ }\textbf {\bibinfo {volume} {54}},\
  \bibinfo {pages} {857} (\bibinfo {year} {1985})}\BibitemShut {NoStop}%
\bibitem [{\citenamefont {Emary}\ \emph {et~al.}(2014)\citenamefont {Emary},
  \citenamefont {Lambert},\ and\ \citenamefont
  {Nori}}]{EmaryLambertNori-sequential}%
  \BibitemOpen
  \bibfield  {author} {\bibinfo {author} {\bibfnamefont {C.}~\bibnamefont
  {Emary}}, \bibinfo {author} {\bibfnamefont {N.}~\bibnamefont {Lambert}},\
  and\ \bibinfo {author} {\bibfnamefont {F.}~\bibnamefont {Nori}},\ }\bibfield
  {title} {\bibinfo {title} {{Leggett-Garg inequalities}},\ }\href
  {https://doi.org/10.1088/0034-4885/77/1/016001} {\bibfield  {journal}
  {\bibinfo  {journal} {Reports on Progress in Physics}\ }\textbf {\bibinfo
  {volume} {77}},\ \bibinfo {pages} {016001} (\bibinfo {year}
  {2014})}\BibitemShut {NoStop}%
\bibitem [{\citenamefont {Clemente}\ and\ \citenamefont
  {Kofler}(2016)}]{ClementeKofler-sequential}%
  \BibitemOpen
  \bibfield  {author} {\bibinfo {author} {\bibfnamefont {L.}~\bibnamefont
  {Clemente}}\ and\ \bibinfo {author} {\bibfnamefont {J.}~\bibnamefont
  {Kofler}},\ }\bibfield  {title} {\bibinfo {title} {{No Fine theorem for
  macrorealism: Limitations of the Leggett-Garg inequality}},\ }\href
  {https://doi.org/10.1103/PhysRevLett.116.150401} {\bibfield  {journal}
  {\bibinfo  {journal} {Physical Review Letters}\ }\textbf {\bibinfo {volume}
  {116}},\ \bibinfo {pages} {150401} (\bibinfo {year} {2016})}\BibitemShut
  {NoStop}%
\bibitem [{\citenamefont {Uola}\ \emph {et~al.}(2019)\citenamefont {Uola},
  \citenamefont {Vitagliano},\ and\ \citenamefont
  {Budroni}}]{UolaVitaglianoBudroni-sequential}%
  \BibitemOpen
  \bibfield  {author} {\bibinfo {author} {\bibfnamefont {R.}~\bibnamefont
  {Uola}}, \bibinfo {author} {\bibfnamefont {G.}~\bibnamefont {Vitagliano}},\
  and\ \bibinfo {author} {\bibfnamefont {C.}~\bibnamefont {Budroni}},\
  }\bibfield  {title} {\bibinfo {title} {{Leggett-Garg macrorealism and the
  quantum nondisturbance conditions}},\ }\href
  {https://doi.org/10.1103/PhysRevA.100.042117} {\bibfield  {journal} {\bibinfo
   {journal} {Physical Review A}\ }\textbf {\bibinfo {volume} {100}},\ \bibinfo
  {pages} {042117} (\bibinfo {year} {2019})}\BibitemShut {NoStop}%
\bibitem [{\citenamefont {Wigner}(1932)}]{Wigner-WignerFunctions}%
  \BibitemOpen
  \bibfield  {author} {\bibinfo {author} {\bibfnamefont {E.}~\bibnamefont
  {Wigner}},\ }\bibfield  {title} {\bibinfo {title} {{On the Quantum Correction
  For Thermodynamic Equilibrium}},\ }\href
  {https://doi.org/10.1103/PhysRev.40.749} {\bibfield  {journal} {\bibinfo
  {journal} {Physical Review}\ }\textbf {\bibinfo {volume} {40}},\ \bibinfo
  {pages} {749} (\bibinfo {year} {1932})}\BibitemShut {NoStop}%
\bibitem [{\citenamefont {Mari}\ and\ \citenamefont
  {Eisert}(2012)}]{MariEisert-WignerFunctionsComputation}%
  \BibitemOpen
  \bibfield  {author} {\bibinfo {author} {\bibfnamefont {A.}~\bibnamefont
  {Mari}}\ and\ \bibinfo {author} {\bibfnamefont {J.}~\bibnamefont {Eisert}},\
  }\bibfield  {title} {\bibinfo {title} {Positive wigner functions render
  classical simulation of quantum computation efficient},\ }\href
  {https://doi.org/10.1103/PhysRevLett.109.230503} {\bibfield  {journal}
  {\bibinfo  {journal} {Physical Review Letters}\ }\textbf {\bibinfo {volume}
  {109}},\ \bibinfo {pages} {230503} (\bibinfo {year} {2012})}\BibitemShut
  {NoStop}%
\bibitem [{\citenamefont {Howard}\ \emph {et~al.}(2014)\citenamefont {Howard},
  \citenamefont {Wallman}, \citenamefont {Veitch},\ and\ \citenamefont
  {Emerson}}]{HowardWallmanVeitchEmerson-ComputationContextuality}%
  \BibitemOpen
  \bibfield  {author} {\bibinfo {author} {\bibfnamefont {M.}~\bibnamefont
  {Howard}}, \bibinfo {author} {\bibfnamefont {J.}~\bibnamefont {Wallman}},
  \bibinfo {author} {\bibfnamefont {V.}~\bibnamefont {Veitch}},\ and\ \bibinfo
  {author} {\bibfnamefont {J.}~\bibnamefont {Emerson}},\ }\bibfield  {title}
  {\bibinfo {title} {Contextuality supplies the ‘magic’ for quantum
  computation},\ }\href {https://doi.org/10.1038/nature13460} {\bibfield
  {journal} {\bibinfo  {journal} {Nature}\ }\textbf {\bibinfo {volume} {510}},\
  \bibinfo {pages} {351} (\bibinfo {year} {2014})}\BibitemShut {NoStop}%
\bibitem [{\citenamefont {Lin}\ and\ \citenamefont
  {Dahlsten}(2020)}]{LinDahlsten-tunneling}%
  \BibitemOpen
  \bibfield  {author} {\bibinfo {author} {\bibfnamefont {Y.~L.}\ \bibnamefont
  {Lin}}\ and\ \bibinfo {author} {\bibfnamefont {O.~C.~O.}\ \bibnamefont
  {Dahlsten}},\ }\bibfield  {title} {\bibinfo {title} {{Necessity of negative
  Wigner function for tunneling}},\ }\href
  {https://doi.org/10.1103/PhysRevA.102.062210} {\bibfield  {journal} {\bibinfo
   {journal} {Physical Review A}\ }\textbf {\bibinfo {volume} {102}},\ \bibinfo
  {pages} {062210} (\bibinfo {year} {2020})}\BibitemShut {NoStop}%
\bibitem [{\citenamefont {Ghai}\ \emph {et~al.}(2023)\citenamefont {Ghai},
  \citenamefont {Sharma},\ and\ \citenamefont
  {Ghosh}}]{GhaiSharmaGhosh-WignerFunctionsIncomaptibility}%
  \BibitemOpen
  \bibfield  {author} {\bibinfo {author} {\bibfnamefont {J.}~\bibnamefont
  {Ghai}}, \bibinfo {author} {\bibfnamefont {G.}~\bibnamefont {Sharma}},\ and\
  \bibinfo {author} {\bibfnamefont {S.}~\bibnamefont {Ghosh}},\ }\bibfield
  {title} {\bibinfo {title} {Negativity of wigner distribution function as a
  measure of incompatibility},\ }\Eprint {https://arxiv.org/abs/2306.07917}
  {arXiv:2306.07917}  (\bibinfo {year} {2023})\BibitemShut {NoStop}%
\bibitem [{\citenamefont {Ohst}\ and\ \citenamefont
  {Pl{\'a}vala}(2023)}]{OhstPlavala-WignerFunctionsSymmetries}%
  \BibitemOpen
  \bibfield  {author} {\bibinfo {author} {\bibfnamefont {T.-A.}\ \bibnamefont
  {Ohst}}\ and\ \bibinfo {author} {\bibfnamefont {M.}~\bibnamefont
  {Pl{\'a}vala}},\ }\bibfield  {title} {\bibinfo {title} {{Symmetries and
  Wigner representations of operational theories}},\ }\Eprint
  {https://arxiv.org/abs/2306.11519} {arXiv:2306.11519}  (\bibinfo {year}
  {2023})\BibitemShut {NoStop}%
\bibitem [{\citenamefont {Tan}\ \emph {et~al.}(2020)\citenamefont {Tan},
  \citenamefont {Choi},\ and\ \citenamefont
  {Jeong}}]{TanChoiJeong-WignerFunctionsNonclassicality}%
  \BibitemOpen
  \bibfield  {author} {\bibinfo {author} {\bibfnamefont {K.~C.}\ \bibnamefont
  {Tan}}, \bibinfo {author} {\bibfnamefont {S.}~\bibnamefont {Choi}},\ and\
  \bibinfo {author} {\bibfnamefont {H.}~\bibnamefont {Jeong}},\ }\bibfield
  {title} {\bibinfo {title} {Negativity of quasiprobability distributions as a
  measure of nonclassicality},\ }\href
  {https://doi.org/10.1103/PhysRevLett.124.110404} {\bibfield  {journal}
  {\bibinfo  {journal} {Physical Review Letters}\ }\textbf {\bibinfo {volume}
  {124}},\ \bibinfo {pages} {110404} (\bibinfo {year} {2020})}\BibitemShut
  {NoStop}%
\bibitem [{\citenamefont {Chu}\ \emph {et~al.}(2018)\citenamefont {Chu},
  \citenamefont {Kharel}, \citenamefont {Yoon}, \citenamefont {Frunzio},
  \citenamefont {Rakich},\ and\ \citenamefont {Schoelkopf}}]{Chu2018}%
  \BibitemOpen
  \bibfield  {author} {\bibinfo {author} {\bibfnamefont {Y.}~\bibnamefont
  {Chu}}, \bibinfo {author} {\bibfnamefont {P.}~\bibnamefont {Kharel}},
  \bibinfo {author} {\bibfnamefont {T.}~\bibnamefont {Yoon}}, \bibinfo {author}
  {\bibfnamefont {L.}~\bibnamefont {Frunzio}}, \bibinfo {author} {\bibfnamefont
  {P.~T.}\ \bibnamefont {Rakich}},\ and\ \bibinfo {author} {\bibfnamefont
  {R.~J.}\ \bibnamefont {Schoelkopf}},\ }\bibfield  {title} {\bibinfo {title}
  {{Creation and control of multi-phonon Fock states in a bulk acoustic-wave
  resonator}},\ }\href {https://doi.org/10.1038/s41586-018-0717-7} {\bibfield
  {journal} {\bibinfo  {journal} {Nature}\ }\textbf {\bibinfo {volume} {563}},\
  \bibinfo {pages} {666} (\bibinfo {year} {2018})}\BibitemShut {NoStop}%
\bibitem [{\citenamefont {Satzinger}\ \emph {et~al.}(2018)\citenamefont
  {Satzinger}, \citenamefont {Zhong}, \citenamefont {Chang}, \citenamefont
  {Peairs}, \citenamefont {Bienfait}, \citenamefont {Chou}, \citenamefont
  {Cleland}, \citenamefont {Conner}, \citenamefont {Dumur}, \citenamefont
  {Grebel}, \citenamefont {Gutierrez}, \citenamefont {November}, \citenamefont
  {Povey}, \citenamefont {Whiteley}, \citenamefont {Awschalom}, \citenamefont
  {Schuster},\ and\ \citenamefont {Cleland}}]{Satzinger18}%
  \BibitemOpen
  \bibfield  {author} {\bibinfo {author} {\bibfnamefont {K.~J.}\ \bibnamefont
  {Satzinger}}, \bibinfo {author} {\bibfnamefont {Y.~P.}\ \bibnamefont
  {Zhong}}, \bibinfo {author} {\bibfnamefont {H.~S.}\ \bibnamefont {Chang}},
  \bibinfo {author} {\bibfnamefont {G.~A.}\ \bibnamefont {Peairs}}, \bibinfo
  {author} {\bibfnamefont {A.}~\bibnamefont {Bienfait}}, \bibinfo {author}
  {\bibfnamefont {M.-H.}\ \bibnamefont {Chou}}, \bibinfo {author}
  {\bibfnamefont {A.~Y.}\ \bibnamefont {Cleland}}, \bibinfo {author}
  {\bibfnamefont {C.~R.}\ \bibnamefont {Conner}}, \bibinfo {author}
  {\bibfnamefont {{\'E}.}~\bibnamefont {Dumur}}, \bibinfo {author}
  {\bibfnamefont {J.}~\bibnamefont {Grebel}}, \bibinfo {author} {\bibfnamefont
  {I.}~\bibnamefont {Gutierrez}}, \bibinfo {author} {\bibfnamefont {B.~H.}\
  \bibnamefont {November}}, \bibinfo {author} {\bibfnamefont {R.~G.}\
  \bibnamefont {Povey}}, \bibinfo {author} {\bibfnamefont {S.~J.}\ \bibnamefont
  {Whiteley}}, \bibinfo {author} {\bibfnamefont {D.~D.}\ \bibnamefont
  {Awschalom}}, \bibinfo {author} {\bibfnamefont {D.~I.}\ \bibnamefont
  {Schuster}},\ and\ \bibinfo {author} {\bibfnamefont {A.~N.}\ \bibnamefont
  {Cleland}},\ }\bibfield  {title} {\bibinfo {title} {Quantum control of
  surface acoustic-wave phonons},\ }\href
  {https://doi.org/10.1038/s41586-018-0719-5} {\bibfield  {journal} {\bibinfo
  {journal} {Nature}\ }\textbf {\bibinfo {volume} {563}},\ \bibinfo {pages}
  {661} (\bibinfo {year} {2018})}\BibitemShut {NoStop}%
\bibitem [{\citenamefont {Wollack}\ \emph {et~al.}(2022)\citenamefont
  {Wollack}, \citenamefont {Cleland}, \citenamefont {Gruenke}, \citenamefont
  {Wang}, \citenamefont {Arrangoiz-Arriola},\ and\ \citenamefont
  {Safavi-Naeini}}]{Wollack22}%
  \BibitemOpen
  \bibfield  {author} {\bibinfo {author} {\bibfnamefont {E.~A.}\ \bibnamefont
  {Wollack}}, \bibinfo {author} {\bibfnamefont {A.~Y.}\ \bibnamefont
  {Cleland}}, \bibinfo {author} {\bibfnamefont {R.~G.}\ \bibnamefont
  {Gruenke}}, \bibinfo {author} {\bibfnamefont {Z.}~\bibnamefont {Wang}},
  \bibinfo {author} {\bibfnamefont {P.}~\bibnamefont {Arrangoiz-Arriola}},\
  and\ \bibinfo {author} {\bibfnamefont {A.~H.}\ \bibnamefont
  {Safavi-Naeini}},\ }\bibfield  {title} {\bibinfo {title} {Quantum state
  preparation and tomography of entangled mechanical resonators},\ }\href
  {https://doi.org/10.1038/s41586-022-04500-y} {\bibfield  {journal} {\bibinfo
  {journal} {Nature}\ }\textbf {\bibinfo {volume} {604}},\ \bibinfo {pages}
  {463} (\bibinfo {year} {2022})}\BibitemShut {NoStop}%
\bibitem [{\citenamefont {von L{\"u}pke}\ \emph {et~al.}(2022)\citenamefont
  {von L{\"u}pke}, \citenamefont {Yang}, \citenamefont {Bild}, \citenamefont
  {Michaud}, \citenamefont {Fadel},\ and\ \citenamefont {Chu}}]{von2022parity}%
  \BibitemOpen
  \bibfield  {author} {\bibinfo {author} {\bibfnamefont {U.}~\bibnamefont {von
  L{\"u}pke}}, \bibinfo {author} {\bibfnamefont {Y.}~\bibnamefont {Yang}},
  \bibinfo {author} {\bibfnamefont {M.}~\bibnamefont {Bild}}, \bibinfo {author}
  {\bibfnamefont {L.}~\bibnamefont {Michaud}}, \bibinfo {author} {\bibfnamefont
  {M.}~\bibnamefont {Fadel}},\ and\ \bibinfo {author} {\bibfnamefont
  {Y.}~\bibnamefont {Chu}},\ }\bibfield  {title} {\bibinfo {title} {Parity
  measurement in the strong dispersive regime of circuit quantum
  acoustodynamics},\ }\href {https://doi.org/10.1038/s41567-022-01591-2}
  {\bibfield  {journal} {\bibinfo  {journal} {Nature Physics}\ }\textbf
  {\bibinfo {volume} {18}},\ \bibinfo {pages} {794} (\bibinfo {year}
  {2022})}\BibitemShut {NoStop}%
\bibitem [{\citenamefont {Bild}\ \emph {et~al.}(2023)\citenamefont {Bild},
  \citenamefont {Fadel}, \citenamefont {Yang}, \citenamefont {von Lüpke},
  \citenamefont {Martin}, \citenamefont {Bruno},\ and\ \citenamefont
  {Chu}}]{bild2022}%
  \BibitemOpen
  \bibfield  {author} {\bibinfo {author} {\bibfnamefont {M.}~\bibnamefont
  {Bild}}, \bibinfo {author} {\bibfnamefont {M.}~\bibnamefont {Fadel}},
  \bibinfo {author} {\bibfnamefont {Y.}~\bibnamefont {Yang}}, \bibinfo {author}
  {\bibfnamefont {U.}~\bibnamefont {von Lüpke}}, \bibinfo {author}
  {\bibfnamefont {P.}~\bibnamefont {Martin}}, \bibinfo {author} {\bibfnamefont
  {A.}~\bibnamefont {Bruno}},\ and\ \bibinfo {author} {\bibfnamefont
  {Y.}~\bibnamefont {Chu}},\ }\bibfield  {title} {\bibinfo {title}
  {Schrödinger cat states of a 16-microgram mechanical oscillator},\ }\href
  {https://doi.org/10.1126/science.adf7553} {\bibfield  {journal} {\bibinfo
  {journal} {Science}\ }\textbf {\bibinfo {volume} {380}},\ \bibinfo {pages}
  {274} (\bibinfo {year} {2023})}\BibitemShut {NoStop}%
\bibitem [{\citenamefont {Tsirelson}(2006)}]{Tsirelson-oscillatorIneq}%
  \BibitemOpen
  \bibfield  {author} {\bibinfo {author} {\bibfnamefont {B.}~\bibnamefont
  {Tsirelson}},\ }\bibfield  {title} {\bibinfo {title} {How often is the
  coordinate of a harmonic oscillator positive?},\ }\Eprint
  {https://arxiv.org/abs/quant-ph/0611147} {arXiv:quant-ph/0611147}  (\bibinfo
  {year} {2006})\BibitemShut {NoStop}%
\bibitem [{\citenamefont {Filip}\ and\ \citenamefont
  {Mišta}(2011)}]{FilipMista-WignerFunctionsDetectingPositive}%
  \BibitemOpen
  \bibfield  {author} {\bibinfo {author} {\bibfnamefont {R.}~\bibnamefont
  {Filip}}\ and\ \bibinfo {author} {\bibfnamefont {L.}~\bibnamefont {Mišta}},\
  }\bibfield  {title} {\bibinfo {title} {Detecting quantum states with a
  positive {W}igner function beyond mixtures of gaussian states},\ }\href
  {https://doi.org/10.1103/PhysRevLett.106.200401} {\bibfield  {journal}
  {\bibinfo  {journal} {Physical Review Letters}\ }\textbf {\bibinfo {volume}
  {106}},\ \bibinfo {pages} {200401} (\bibinfo {year} {2011})}\BibitemShut
  {NoStop}%
\bibitem [{\citenamefont {Chabaud}\ \emph {et~al.}(2021)\citenamefont
  {Chabaud}, \citenamefont {Emeriau},\ and\ \citenamefont
  {Grosshans}}]{ChabaudEmeriauGrosshans-WignerFunctionNegativeWitness}%
  \BibitemOpen
  \bibfield  {author} {\bibinfo {author} {\bibfnamefont {U.}~\bibnamefont
  {Chabaud}}, \bibinfo {author} {\bibfnamefont {P.-E.}\ \bibnamefont
  {Emeriau}},\ and\ \bibinfo {author} {\bibfnamefont {F.}~\bibnamefont
  {Grosshans}},\ }\bibfield  {title} {\bibinfo {title} {Witnessing {W}igner
  negativity},\ }\href {https://doi.org/10.22331/q-2021-06-08-471} {\bibfield
  {journal} {\bibinfo  {journal} {Quantum}\ }\textbf {\bibinfo {volume} {5}},\
  \bibinfo {pages} {471} (\bibinfo {year} {2021})}\BibitemShut {NoStop}%
\bibitem [{\citenamefont {Zaw}\ \emph {et~al.}(2022)\citenamefont {Zaw},
  \citenamefont {Aw}, \citenamefont {Lasmar},\ and\ \citenamefont
  {Scarani}}]{ZawCenxinLasmarScarani-spinTsirelsonIneq}%
  \BibitemOpen
  \bibfield  {author} {\bibinfo {author} {\bibfnamefont {L.~H.}\ \bibnamefont
  {Zaw}}, \bibinfo {author} {\bibfnamefont {C.~C.}\ \bibnamefont {Aw}},
  \bibinfo {author} {\bibfnamefont {Z.}~\bibnamefont {Lasmar}},\ and\ \bibinfo
  {author} {\bibfnamefont {V.}~\bibnamefont {Scarani}},\ }\bibfield  {title}
  {\bibinfo {title} {Detecting quantumness in uniform precessions},\ }\href
  {https://doi.org/10.1103/PhysRevA.106.032222} {\bibfield  {journal} {\bibinfo
   {journal} {Physical Review A}\ }\textbf {\bibinfo {volume} {106}},\ \bibinfo
  {pages} {032222} (\bibinfo {year} {2022})}\BibitemShut {NoStop}%
\bibitem [{\citenamefont {Zaw}\ and\ \citenamefont
  {Scarani}(2023)}]{ZawScarani-anharmonicTsirelsonIneq}%
  \BibitemOpen
  \bibfield  {author} {\bibinfo {author} {\bibfnamefont {L.~H.}\ \bibnamefont
  {Zaw}}\ and\ \bibinfo {author} {\bibfnamefont {V.}~\bibnamefont {Scarani}},\
  }\bibfield  {title} {\bibinfo {title} {Dynamics-based quantumness
  certification of continuous variables using time-independent hamiltonians
  with one degree of freedom},\ }\href
  {https://doi.org/10.1103/PhysRevA.108.022211} {\bibfield  {journal} {\bibinfo
   {journal} {Physical Review A}\ }\textbf {\bibinfo {volume} {108}},\ \bibinfo
  {pages} {022211} (\bibinfo {year} {2023})}\BibitemShut {NoStop}%
\bibitem [{\citenamefont {Jayachandran}\ \emph {et~al.}(2023)\citenamefont
  {Jayachandran}, \citenamefont {Zaw},\ and\ \citenamefont
  {Scarani}}]{JayachandranZawScarani-entanglementTsirelsonIneq}%
  \BibitemOpen
  \bibfield  {author} {\bibinfo {author} {\bibfnamefont {P.}~\bibnamefont
  {Jayachandran}}, \bibinfo {author} {\bibfnamefont {L.~H.}\ \bibnamefont
  {Zaw}},\ and\ \bibinfo {author} {\bibfnamefont {V.}~\bibnamefont {Scarani}},\
  }\bibfield  {title} {\bibinfo {title} {Dynamics-based entanglement witnesses
  for non-gaussian states of harmonic oscillators},\ }\href
  {https://doi.org/10.1103/PhysRevLett.130.160201} {\bibfield  {journal}
  {\bibinfo  {journal} {Physical Review Letters}\ }\textbf {\bibinfo {volume}
  {130}},\ \bibinfo {pages} {160201} (\bibinfo {year} {2023})}\BibitemShut
  {NoStop}%
\bibitem [{\citenamefont {Gawrilow}\ and\ \citenamefont
  {Joswig}(2000)}]{polymake}%
  \BibitemOpen
  \bibfield  {author} {\bibinfo {author} {\bibfnamefont {E.}~\bibnamefont
  {Gawrilow}}\ and\ \bibinfo {author} {\bibfnamefont {M.}~\bibnamefont
  {Joswig}},\ }\bibfield  {title} {\bibinfo {title} {Polymake: a framework for
  analyzing convex polytopes},\ }in\ \href@noop {} {\emph {\bibinfo {booktitle}
  {Polytopes—combinatorics and computation}}}\ (\bibinfo {organization}
  {Springer},\ \bibinfo {year} {2000})\ pp.\ \bibinfo {pages}
  {43--73}\BibitemShut {NoStop}%
\bibitem [{\citenamefont {Augusiak}\ and\ \citenamefont
  {Stasi{\'n}ska}(2009)}]{augusiak2009positive}%
  \BibitemOpen
  \bibfield  {author} {\bibinfo {author} {\bibfnamefont {R.}~\bibnamefont
  {Augusiak}}\ and\ \bibinfo {author} {\bibfnamefont {J.}~\bibnamefont
  {Stasi{\'n}ska}},\ }\bibfield  {title} {\bibinfo {title} {Positive maps,
  majorization, entropic inequalities and detection of entanglement},\ }\href
  {https://doi.org/10.1088/1367-2630/11/5/053018} {\bibfield  {journal}
  {\bibinfo  {journal} {New Journal of Physics}\ }\textbf {\bibinfo {volume}
  {11}},\ \bibinfo {pages} {053018} (\bibinfo {year} {2009})}\BibitemShut
  {NoStop}%
\bibitem [{\citenamefont {Liang}\ \emph {et~al.}(2011)\citenamefont {Liang},
  \citenamefont {Spekkens},\ and\ \citenamefont
  {Wiseman}}]{LiangSpekkensWiseman-Specker}%
  \BibitemOpen
  \bibfield  {author} {\bibinfo {author} {\bibfnamefont {Y.-C.}\ \bibnamefont
  {Liang}}, \bibinfo {author} {\bibfnamefont {R.~W.}\ \bibnamefont
  {Spekkens}},\ and\ \bibinfo {author} {\bibfnamefont {H.~M.}\ \bibnamefont
  {Wiseman}},\ }\bibfield  {title} {\bibinfo {title} {{Specker’s parable of
  the overprotective seer: A road to contextuality, nonlocality and
  complementarity}},\ }\href {https://doi.org/10.1016/j.physrep.2011.05.001}
  {\bibfield  {journal} {\bibinfo  {journal} {Physics Reports}\ }\textbf
  {\bibinfo {volume} {506}},\ \bibinfo {pages} {1} (\bibinfo {year}
  {2011})}\BibitemShut {NoStop}%
\bibitem [{\citenamefont {Liang}\ \emph {et~al.}(2017)\citenamefont {Liang},
  \citenamefont {Spekkens},\ and\ \citenamefont
  {Wiseman}}]{LiangSpekkensWiseman-SpeckerErratum}%
  \BibitemOpen
  \bibfield  {author} {\bibinfo {author} {\bibfnamefont {Y.~C.}\ \bibnamefont
  {Liang}}, \bibinfo {author} {\bibfnamefont {R.~W.}\ \bibnamefont
  {Spekkens}},\ and\ \bibinfo {author} {\bibfnamefont {H.~M.}\ \bibnamefont
  {Wiseman}},\ }\bibfield  {title} {\bibinfo {title} {{Erratum to “Specker's
  parable of the over-protective seer: A road to contextuality, nonlocality and
  complementarity”}},\ }\href {https://doi.org/10.1016/j.physrep.2016.12.001}
  {\bibfield  {journal} {\bibinfo  {journal} {Physics Reports}\ }\textbf
  {\bibinfo {volume} {666}},\ \bibinfo {pages} {110} (\bibinfo {year}
  {2017})}\BibitemShut {NoStop}%
\bibitem [{\citenamefont {Clauser}\ \emph {et~al.}(1969)\citenamefont
  {Clauser}, \citenamefont {Horne}, \citenamefont {Shimony},\ and\
  \citenamefont {Holt}}]{ClauserHorneShimonyHolt-CHSH}%
  \BibitemOpen
  \bibfield  {author} {\bibinfo {author} {\bibfnamefont {J.~F.}\ \bibnamefont
  {Clauser}}, \bibinfo {author} {\bibfnamefont {M.~A.}\ \bibnamefont {Horne}},
  \bibinfo {author} {\bibfnamefont {A.}~\bibnamefont {Shimony}},\ and\ \bibinfo
  {author} {\bibfnamefont {R.~A.}\ \bibnamefont {Holt}},\ }\bibfield  {title}
  {\bibinfo {title} {{Proposed Experiment to Test Local Hidden-Variable
  Theories}},\ }\href {https://doi.org/10.1103/PhysRevLett.23.880} {\bibfield
  {journal} {\bibinfo  {journal} {Physical Review Letters}\ }\textbf {\bibinfo
  {volume} {23}},\ \bibinfo {pages} {880} (\bibinfo {year} {1969})}\BibitemShut
  {NoStop}%
\bibitem [{\citenamefont {Clauser}\ \emph {et~al.}(1970)\citenamefont
  {Clauser}, \citenamefont {Horne}, \citenamefont {Shimony},\ and\
  \citenamefont {Holt}}]{ClauserHorneShimonyHolt-CHSHerratum}%
  \BibitemOpen
  \bibfield  {author} {\bibinfo {author} {\bibfnamefont {J.~F.}\ \bibnamefont
  {Clauser}}, \bibinfo {author} {\bibfnamefont {M.~A.}\ \bibnamefont {Horne}},
  \bibinfo {author} {\bibfnamefont {A.}~\bibnamefont {Shimony}},\ and\ \bibinfo
  {author} {\bibfnamefont {R.~A.}\ \bibnamefont {Holt}},\ }\bibfield  {title}
  {\bibinfo {title} {{Proposed Experiment to Test Local Hidden Variable
  Theories}},\ }\href {https://doi.org/10.1103/PhysRevLett.24.549} {\bibfield
  {journal} {\bibinfo  {journal} {Physical Review Letters}\ }\textbf {\bibinfo
  {volume} {24}},\ \bibinfo {pages} {549} (\bibinfo {year} {1970})}\BibitemShut
  {NoStop}%
\bibitem [{\citenamefont {Abramsky}\ and\ \citenamefont
  {Brandenburger}(2011)}]{AbramskyBrandenburger-contextuality}%
  \BibitemOpen
  \bibfield  {author} {\bibinfo {author} {\bibfnamefont {S.}~\bibnamefont
  {Abramsky}}\ and\ \bibinfo {author} {\bibfnamefont {A.}~\bibnamefont
  {Brandenburger}},\ }\bibfield  {title} {\bibinfo {title} {The sheaf-theoretic
  structure of non-locality and contextuality},\ }\href
  {https://doi.org/10.1088/1367-2630/13/11/113036} {\bibfield  {journal}
  {\bibinfo  {journal} {New Journal of Physics}\ }\textbf {\bibinfo {volume}
  {13}},\ \bibinfo {pages} {113036} (\bibinfo {year} {2011})}\BibitemShut
  {NoStop}%
\bibitem [{\citenamefont {Abramsky}\ and\ \citenamefont
  {Hardy}(2012)}]{AbramskyHardy-logicalBell}%
  \BibitemOpen
  \bibfield  {author} {\bibinfo {author} {\bibfnamefont {S.}~\bibnamefont
  {Abramsky}}\ and\ \bibinfo {author} {\bibfnamefont {L.}~\bibnamefont
  {Hardy}},\ }\bibfield  {title} {\bibinfo {title} {Logical {B}ell
  inequalities},\ }\href {https://doi.org/10.1103/PhysRevA.85.062114}
  {\bibfield  {journal} {\bibinfo  {journal} {Physical Review A}\ }\textbf
  {\bibinfo {volume} {85}},\ \bibinfo {pages} {062114} (\bibinfo {year}
  {2012})}\BibitemShut {NoStop}%
\bibitem [{\citenamefont {Groenewold}(1946)}]{Groenewold-QM}%
  \BibitemOpen
  \bibfield  {author} {\bibinfo {author} {\bibfnamefont {H.}~\bibnamefont
  {Groenewold}},\ }\bibfield  {title} {\bibinfo {title} {{On the principles of
  elementary quantum mechanics}},\ }\href
  {https://doi.org/10.1016/S0031-8914(46)80059-4} {\bibfield  {journal}
  {\bibinfo  {journal} {Physica}\ }\textbf {\bibinfo {volume} {12}},\ \bibinfo
  {pages} {405} (\bibinfo {year} {1946})}\BibitemShut {NoStop}%
\bibitem [{\citenamefont {Moyal}(1949)}]{Moyal-WignerFunctions}%
  \BibitemOpen
  \bibfield  {author} {\bibinfo {author} {\bibfnamefont {J.~E.}\ \bibnamefont
  {Moyal}},\ }\bibfield  {title} {\bibinfo {title} {{Quantum mechanics as a
  statistical theory}},\ }\href {https://doi.org/10.1017/S0305004100000487}
  {\bibfield  {journal} {\bibinfo  {journal} {Mathematical Proceedings of the
  Cambridge Philosophical Society}\ }\textbf {\bibinfo {volume} {45}},\
  \bibinfo {pages} {99} (\bibinfo {year} {1949})}\BibitemShut {NoStop}%
\bibitem [{\citenamefont {Curtright}\ \emph {et~al.}(2013)\citenamefont
  {Curtright}, \citenamefont {Fairlie},\ and\ \citenamefont
  {Zachos}}]{CurtrightFarlieZachos-WignerFunctions}%
  \BibitemOpen
  \bibfield  {author} {\bibinfo {author} {\bibfnamefont {T.}~\bibnamefont
  {Curtright}}, \bibinfo {author} {\bibfnamefont {D.}~\bibnamefont {Fairlie}},\
  and\ \bibinfo {author} {\bibfnamefont {C.}~\bibnamefont {Zachos}},\
  }\href@noop {} {\emph {\bibinfo {title} {A Concise Treatise On Quantum
  Mechanics In Phase Space}}}\ (\bibinfo  {publisher} {World Scientific
  Publishing Company},\ \bibinfo {year} {2013})\BibitemShut {NoStop}%
\bibitem [{\citenamefont {Case}(2008)}]{Case-wignerFunctions}%
  \BibitemOpen
  \bibfield  {author} {\bibinfo {author} {\bibfnamefont {W.~B.}\ \bibnamefont
  {Case}},\ }\bibfield  {title} {\bibinfo {title} {{Wigner functions and Weyl
  transforms for pedestrians}},\ }\href {https://doi.org/10.1119/1.2957889}
  {\bibfield  {journal} {\bibinfo  {journal} {American Journal of Physics}\
  }\textbf {\bibinfo {volume} {76}},\ \bibinfo {pages} {937} (\bibinfo {year}
  {2008})}\BibitemShut {NoStop}%
\bibitem [{\citenamefont {Allcock}(1969)}]{Allcock-backflow}%
  \BibitemOpen
  \bibfield  {author} {\bibinfo {author} {\bibfnamefont {G.~R.}\ \bibnamefont
  {Allcock}},\ }\bibfield  {title} {\bibinfo {title} {The time of arrival in
  quantum mechanics ii. the individual measurement},\ }\href
  {https://doi.org/10.1016/0003-4916(69)90252-8} {\bibfield  {journal}
  {\bibinfo  {journal} {Annals of Physics}\ }\textbf {\bibinfo {volume} {53}},\
  \bibinfo {pages} {286} (\bibinfo {year} {1969})}\BibitemShut {NoStop}%
\bibitem [{\citenamefont {Bracken}\ and\ \citenamefont
  {Melloy}(1994)}]{Bracken-backflow}%
  \BibitemOpen
  \bibfield  {author} {\bibinfo {author} {\bibfnamefont {A.}~\bibnamefont
  {Bracken}}\ and\ \bibinfo {author} {\bibfnamefont {G.}~\bibnamefont
  {Melloy}},\ }\bibfield  {title} {\bibinfo {title} {Probability backflow and a
  new dimensionless quantum number},\ }\href
  {https://doi.org/10.1088/0305-4470/27/6/040} {\bibfield  {journal} {\bibinfo
  {journal} {Journal of Physics A: Mathematical and General}\ }\textbf
  {\bibinfo {volume} {27}},\ \bibinfo {pages} {2197} (\bibinfo {year}
  {1994})}\BibitemShut {NoStop}%
\bibitem [{\citenamefont {Goussev}(2020)}]{Goussev-reentry}%
  \BibitemOpen
  \bibfield  {author} {\bibinfo {author} {\bibfnamefont {A.}~\bibnamefont
  {Goussev}},\ }\bibfield  {title} {\bibinfo {title} {Probability backflow for
  correlated quantum states},\ }\href
  {https://doi.org/10.1103/PhysRevResearch.2.033206} {\bibfield  {journal}
  {\bibinfo  {journal} {Physical Review Research}\ }\textbf {\bibinfo {volume}
  {2}},\ \bibinfo {pages} {033206} (\bibinfo {year} {2020})}\BibitemShut
  {NoStop}%
\bibitem [{\citenamefont {Trillo}\ \emph {et~al.}(2023)\citenamefont {Trillo},
  \citenamefont {Le},\ and\ \citenamefont
  {Navascu{\'e}s}}]{TrilloLeNavascues-rocket}%
  \BibitemOpen
  \bibfield  {author} {\bibinfo {author} {\bibfnamefont {D.}~\bibnamefont
  {Trillo}}, \bibinfo {author} {\bibfnamefont {T.~P.}\ \bibnamefont {Le}},\
  and\ \bibinfo {author} {\bibfnamefont {M.}~\bibnamefont {Navascu{\'e}s}},\
  }\bibfield  {title} {\bibinfo {title} {Quantum advantages for transportation
  tasks-projectiles, rockets and quantum backflow},\ }\href
  {https://doi.org/10.1038/s41534-023-00739-z} {\bibfield  {journal} {\bibinfo
  {journal} {npj Quantum Information}\ }\textbf {\bibinfo {volume} {9}},\
  \bibinfo {pages} {69} (\bibinfo {year} {2023})}\BibitemShut {NoStop}%
\end{thebibliography}%

\end{document}